\documentclass[twoside]{article}
\usepackage[latin1]{inputenc}
\usepackage{t1enc}
\usepackage{a4}
\usepackage{tabularx}
\usepackage{epsf}
\usepackage{epsfig}

\def\bref{\vspace{4pt}\noindent\hangindent=10mm}

\begin{document}

\setcounter{figure}{0}
\setcounter{section}{0}
\setcounter{equation}{0}

\begin{center}
{\Large\bf
The high mass end of extragalactic\\[0.2cm]
globular clusters}\\[0.7cm]

Michael Hilker\\[0.17cm]
ESO\\
Karl-Schwarzschild-Str. 2\\
mhilker@eso.org, http://www.eso.org/$\sim$mhilker/
\end{center}

\vspace{0.5cm}

\begin{abstract}
\noindent{\it
In the last decade, a new kind of stellar systems has been established that
shows properties in between those of globular clusters (GCs) and early-type 
dwarf galaxies. These so-called ultra-compact dwarf galaxies (UCDs) have masses
in the range $10^6$ to $10^8$ $M_{\odot}$ and half-light radii of 10-100 pc. 
The most massive UCDs known to date are predominantly metal-rich and reside 
in the cores of nearby galaxy clusters. The question arises whether UCDs are 
just the most massive globular clusters in rich globular cluster systems?
Although UCDs and `normal' GCs form a continuous sequence in several parameter
spaces, there seems to be a break in the scaling laws for stellar systems with
masses above $\sim2.5\times10^6$ $M_{\odot}$. Unlike GCs, UCDs follow a 
mass-size relation and their mass-to-light ratios are about twice as large as 
those of GCs with comparable metallicities. In this contribution, I present 
the properties of the brightest globular clusters and ultra-compact 
dwarf galaxies and discuss whether the observed findings are compatible 
with a `star-cluster' origin of UCDs or whether they are more likely related 
to dark matter dominated dwarf galaxies.}
\end{abstract}

\section{The most massive globular clusters of a galaxy}

$\omega$ Centauri is the most luminous and massive globular cluster of our
Galaxy. With an absolute magnitude of $M_V=-10.29$ mag
(Harris 1996) and a mass of $2.5\times10^6 M_{\odot}$ (van de Ven et al. 
2006), it is an order of magnitude more luminous and massive than an average 
Galactic globular cluster ($M_V=-7.5$, $2\times10^5 M_{\odot}$). But can 
$\omega$ Cen actually be regarded as a globular cluster? 
Several studies over the past decade have shown that $\omega$ Cen is composed
of multiple stellar populations with different, rather discrete abundance 
patterns and probably a spread in their ages (e.g. Hilker \& Richtler 2000,
Bedin et al. 2004, Sollima et al. 2005, Villanova et al. 2007). Such a complex
behaviour is usually only seen in galaxies, like the Local Group dwarf 
spheroidals (for example the Carina dSph: Koch et al. 2007).

The view of globular clusters (GCs) as simple stellar systems was even more
revolutionised by studies based on precise HST-based photometry that
revealed multiple stellar populations in several massive Galactic globular 
clusters (e.g. Piotto et al. 2007, Milone et al. 2008). But this is the 
story of another review in this book (see the contribution by Piotto).
Here I concentrate on the properties of the most massive globular clusters 
in external galaxies, and even more massive compact stellar systems in galaxy
clusters. 

Departing from the Milky Way we can first ask what are the properties of 
the most massive globular clusters in other Local Group galaxies? 

The Andromeda galaxy has a $\sim$3 times larger globular cluster system (GCS)
than our Galaxy (e.g. Barmby et al. 2001) and possesses several GCs that 
are $\sim$3 times more luminous/massive than $\omega$ Cen. In particular G1,
one of the most massive clusters in M31, exhibits a spread in its red giant 
branch, probably caused by multiple stellar populations of different 
metallicities (Meylan et al. 2001). At the lower mass end of Local Group 
galaxies, old GCs ($>5$ Gyr) are known in the LMC and SMC (LMC: Mackey \& 
Gilmore 2004; SMC: Crowl et al. 2001, Glatt et al.
2008), the dwarf ellipticals NGC\,205, NGC\,185 and NGC\,147 (Hodge 1993, 1974,
1976; Da Costa \& Mould 1988), and the Fornax and Sagittarius dwarf (Sgr) 
spheroidals (For dSph: Buonanno et al. 1999, Mackey \& Gilmore 2003; Sgr dSph:
Carraro et al. 2007; Carraro 2009). The most luminous GCs in these galaxies 
are 2-3 magnitudes fainter than those in the Milky Way and Andromeda (see 
Table\,1 and Fig.\,2). 

Going to denser environments and more massive galaxies beyond the Local Group
we can then ask whether the trend of more luminous/massive GCs in ever more
luminous galaxies continues or whether there exists some kind of cut-off mass
for the most massive GC? How massive can a GC get?

Finding the most massive GC in distant galaxies is not an easy task. Since
distant GCs are not resolved on ground based images, contamination by 
foreground
stars and compact background galaxies hampers the exact definition of
the sparsely sampled bright end of the globular cluster luminosity function 
(GCLF). Only massive spectroscopic surveys and the resolved appearance of 
GCs on HST images made it possible to discover the brightest GCs at distances
beyond the Local Group. In this respect, the best studied GCSs of nearby 
elliptical galaxies are those of Centaurus\,A (e.g. Peng et al. 2004, Rejkuba 
et al. 2007), 
NGC\,1399 (Drinkwater et al. 2000, Mieske et al. 2004) and M\,87 (Ha\c{s}egan 
et al. 2005, Jones et al. 2006), the central galaxies 
of the Centaurus group, the Fornax and the Virgo cluster, respectively.
Indeed, compact sources with masses up to a hundred times that of 
$\omega$\,Cen have been identified. Their discovery history and properties 
are described in the next section.

\begin{table}[h!]
\caption{Properties of brightest GCs and UCDs and their host galaxies.}
\begin{tabular}{lrrrlrr}
 & & & & & & \\[-3mm]
\hline\noalign{\smallskip}
Galaxy & $M_{V,\rm gal}$ & $N_{\rm GC,tot}$ & $\sigma_{\rm GCLF}$ & GC Name &
$M_{V,\rm GC}$ & $\log$(M) \\
 & [mag] & & [mag] & & [mag] & M$_{\odot}$ \\[1mm]
\hline\noalign{\smallskip}
Fnx dSph & $-$13.1 & 5 & 0.50 & Fnx3 & $-$7.80 & 5.560 \\
 & & & & Fnx2 & $-$7.05 & 5.260 \\
\noalign{\smallskip}
Sgr dSph & $-$15.0 & 7 & 0.60 & M54 & $-$8.55 & 5.857 \\
 & & & & Arp\,2 & $-$5.60 & 4.040 \\
\noalign{\smallskip}
NGC\,147 & $-$15.1 & 4 & 0.60 & NGC\,147-3 & $-$7.93 & 5.484 \\
 & & & & NGC\,147-1 & $-$7.23 & 5.222 \\
\noalign{\smallskip}
NGC\,185 & $-$15.6 & 8 & 0.65 & NGC\,185-5 & $-$7.83 & 5.479 \\
 & & & & NGC\,185-3 & $-$7.73 & 5.447 \\
\noalign{\smallskip}
NGC\,205 & $-$16.4 & 8 & 0.70 & NGC\,205-8 & $-$8.19 & 5.606 \\
 & & & & NGC\,205-2 & $-$8.09 & 5.599 \\
\noalign{\smallskip}
SMC & $-$17.1 & 8 & 0.80 & Kron\,3 & $-$8.00 & 5.350 \\
 & & & & NGC\,121 & $-$7.94 & 5.550 \\
 & & & & NGC\,416 & $-$7.70 & 5.270 \\
\noalign{\smallskip}
LMC & $-$18.5 & 16 & 0.90 & NGC\,1898 & $-$8.60 & 5.880 \\
 & & & & NGC\,1835 & $-$8.33 & 5.830 \\
 & & & & NGC\,1916 & $-$8.33 & 5.790 \\
\noalign{\smallskip}
Milky Way & $-$20.9 & 150 & 1.15 & $\omega$\,Cen & $-$10.29 & 6.398 \\
 & & & & NGC\,6715 & $-$10.01 & 6.240 \\
 & & & & NGC\,6441 & $-$9.64 & 6.170 \\
\noalign{\smallskip}
M31 & $-$21.2 & 460 & 1.20 & B023 & $-$11.33 & 6.955 \\
 & & & & G1 & $-$10.94 & 6.863 \\
 & & & & B225 & $-$10.75 & 6.778 \\
\noalign{\smallskip}
Cen\,A & $-$21.5 & 1550 & 1.30 & HCH99-18 & $-$11.38 & 7.050 \\
 & & & & HGHH92-C1 & $-$10.84 & 6.833 \\
 & & & & HGHH92-C23 & $-$11.66 & 6.822 \\
\noalign{\smallskip}
NGC\,1399 & $-$21.9 & 6450 & 1.25 & UCD3 & $-$13.40 & 7.971 \\
 & & & & UCD1 & $-$12.07 & 7.507 \\
 & & & & UCD6 & $-$12.50 & 7.476 \\
\noalign{\smallskip}
M87 & $-$22.4 & 14660 & 1.30 & VUCD7 & $-$13.42 & 7.946 \\
 & & & & VUCD3 & $-$12.59 & 7.602 \\
 & & & & VUCD5 & $-$12.32 & 7.464 \\
\noalign{\smallskip}\hline
\end{tabular}
\end{table}

\section{Ultra-Compact Dwarf Galaxies}

The discovery history of very massive compact objects started about
10 years ago. In a small spectroscopic survey of the globular cluster system
of NGC\,1399, Minniti et al. (1998) confirmed a bright compact object as 
radial velocity member of the cluster: {\it `... Note that the object at 
$V=18.5$, $V-I=1.48$ (our reddest ``globular cluster''), which has $M_V = 
-12.5$, was identified as a compact dwarf galaxy on the images after 
light-profile analysis (M. Hilker, 1996, private communication) ...''} (see 
also Hilker 1998). In another spectroscopic survey on dwarf ellipticals in the 
Fornax cluster, Hilker et al. (1999) confirmed two bright compact objects 
with $M_V=-13.4$ and $-12.6$ mag (including the one mentioned before) as 
Fornax members. They proposed that they {\it `... can be explained by a very 
bright GC as well as by a compact elliptical like M32. Another explanation 
might be that these objects represent the nuclei of dissolved dE,Ns ...'}. 
Furthermore they suggested that {\it `... It 
would be interesting to investigate, whether there are more objects of this 
kind hidden among the high surface brightness objects in the central Fornax 
cluster ...'}.

Indeed, only one year later, in 2000, a systematic all-object spectroscopic 
survey within in a 2-degree field centred on the Fornax cluster revealed five 
compact Fornax members in the magnitude range $-13.5<M_V<-12.0$ (Drinkwater et
al. 2000) which later, in 2001, were dubbed ``Ultracompact Dwarf Galaxies'' 
(UCDs) by Phillipps et al. (2001). Their physical properties were presented in 
a {\it Nature} article by Drinkwater et al. (2003). Later, Mieske et al. 
(2004) identified compact objects in the brightness range $-12.0<M_V<-10.0$ 
mag. They found that their luminosity distribution is consistent with an 
extrapolation of the Gaussian-shaped GC luminosity function.

After the first discovery of UCDs in the Fornax cluster, many surveys followed
to search for UCDs in different environments and towards fainter magnitudes 
(Virgo cluster (M87): Ha\c{s}egan et al. 2005, Jones at al. 2006; Centaurus 
cluster (NGC\,4696): Mieske et al. 2007; Hydra\,I cluster (NGC\,3311): Misgeld
et al. 2008; Cen\,A: Rejkuba et al. 2007; Sombrero: Hau et al. 2009). 
Although massive UCDs mainly are found in galaxy clusters and therefore might 
be linked to the overall cluster formation process, most of them seem to be
associated to giant galaxies. Regarding their radial distribution
and kinematic signature around their host galaxies, UCDs can hardly be 
distinguished from luminous/massive genuine globular clusters belonging to 
those galaxies. Therefore, I consider objects more luminous than $M_V<-11$ 
mag -- GCs as well as UCDs -- as
one class and simply call them `UCDs' or sometimes `GCs/UCDs' throughout this
contribution, being aware of the fact that the formation processes of UCDs in 
the cluster environment and massive GCs around individual galaxies might be
different.

Once the existence of UCDs was proven by radial velocity measurements, further
studies focused on their physical parameters. In 
particular, their sizes, metallicities, ages, internal kinematics, masses and 
mass-to-light ratios were investigated. The most important results are 
summarized in the following.

\begin{figure}[t!]
\epsfysize=10.0cm
\epsfbox{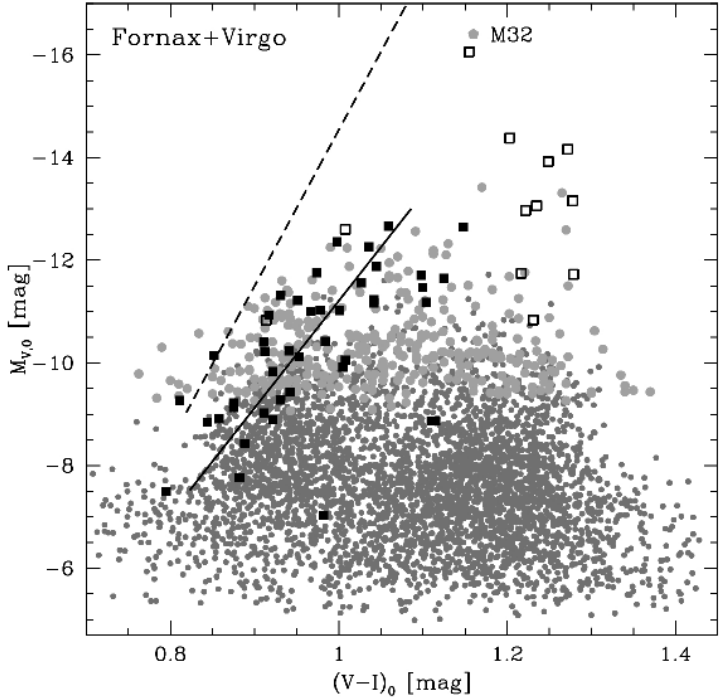}
\vspace*{-0.2cm}
\caption{Colour magnitude diagram of GCs, UCDs and nuclear clusters in the
Fornax and Virgo clusters. Small grey dots represent GCs around NGC\,1399 and
NGC\,1404 (Jord\'an et al. 2009) and M\,87 and M\,49 (Peng et al. 2006) from 
HST/ACS data. Large grey dots are confirmed cluster members (Hilker 2009, in 
prep.) and filled and open squares mark the nuclei of early-ype galaxies in 
Virgo (C\^ot\'e et al. 2006). The solid line is a fit to the filled squares, 
whereas the dashed line represents the colour-magnitude relation of dEs in 
Fornax (Mieske et al. 2007b). The location of M32 is shown as well.
}
\end{figure}

UCDs are luminous ($-11.0<M_V<-13.5$), have half-light radii in the range
$10<r_{\rm h}<100$ pc and are predominantly old ($>10$ Gyr) (e.g. Mieske et 
al. 2006; Evstigneeva et al. 2007). As opposed to GCs, UCDs follow a
luminosity-size relation (e.g. Ha\c{s}egan et al. 2005; Evstigneeva et al. 
2008). M32-type galaxies lie on the extension of this relation (Dabringhausen
et al. 2008). Also nuclei of early-type galaxies exhibit a luminosity-size 
relation, shifted towards smaller sizes at a given luminosity (C\^ot\'e et al.
2006). The two brightest UCDs in Fornax (UCD3) and Virgo (VUCD7), both with 
$M_V\simeq-13.5$, are at least twice as luminous as the second brightest UCD 
in their 
respective clusters. They exhibit faint surface brightness envelopes with 
effective radii of $80<R_{\rm eff}<120$ pc (Evstigneeva et al. 2007). 

In the colour-magnitude diagram (see Fig.\,1), UCDs cover the full colour 
range of `normal' GCs. However, the brightest UCDs are found on the extension
of the red (metal-rich) GC population (Mieske et al. 2006; Wehner \& Harris
2008). Blue (metal-poor) UCDs coincide with the location of nuclear clusters
in early-type dwarf galaxies.

The central velocity dispersions of UCDs range from 15 to 45 km~s$^{-1}$,
resulting in dynamical masses of $2\times10^6<M<10^8 M_\odot$ (e.g. Hilker et
al. 2007, Mieske et al. 2008). The most remarkable consequence of these
derived masses is that the dynamical mass-to-light ratio of UCDs is on
average twice that of GCs at comparable metallicity and cannot
be explained by stellar population models with a canonical initial mass
function (IMF, e.g. Kroupa 2001) (Ha\c{s}egan et al. 2005, Dabringhausen et
al. 2008, Mieske et al. 2008). The large $M/L$ values of UCDs might either be
caused by an unusual IMF (bottom-heavy: Mieske \& Kroupa 2008; top-heavy:
Dabringhausen et al. 2009) or by the presence of dark matter (Baumgardt \&
Mieske 2008).

All the properties presented above and the scaling relations of UCDs hint to
a characteristic transition mass of $M_c\simeq 2.5\times10^6 M_\odot$ between
GCs and UCDs. This does not necessarily mean that GCs and UCDs are different
kinds of objects. It might just reflect a change in the physics of cluster
formation at this characteristic mass, for example, if more massive clusters
become optically thick to far infrared radiation when they formed and are born
with top-heavy IMFs (Murray 2009).

\begin{figure}[t!]
\epsfysize=10.0cm
\epsfbox{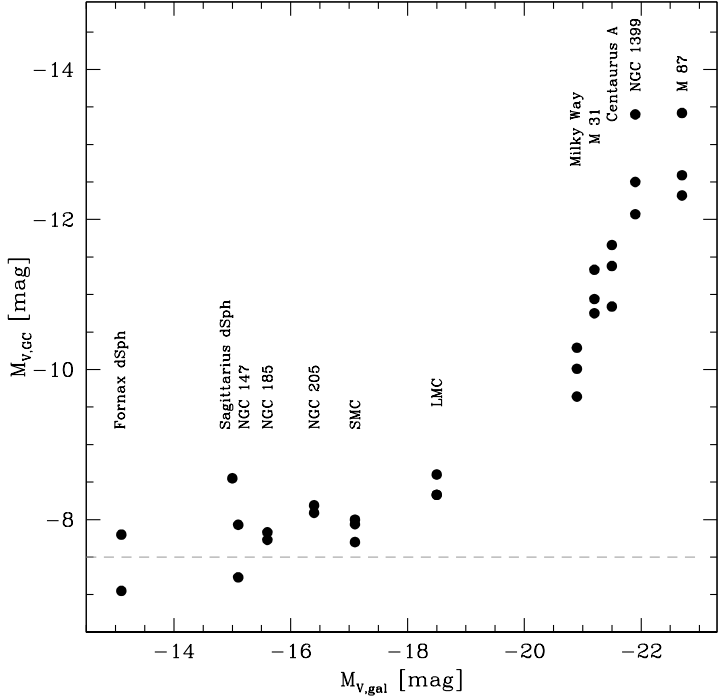}
\vspace*{-0.2cm}
\caption{The absolute magnitude of the brightest two or three GCs/UCDs of
a galaxy as a function of host galaxy luminosity. The dashed line indicates
the universal luminosity of the GCLF turnover magnitude.
}
\end{figure}

In the next section we will investigate whether the transition from GCs to
UCDs can be seen in the luminosity and mass function of well studied
globular cluster systems and UCD populations.

\section{Luminosity and mass function of GCs/UCDs}

In Fig.\,2 the luminosities of the two or three brightest GCs (and UCDs)
are plotted as function of host galaxy luminosity for all the galaxies
discussed in Sect.\,1 (see the parameters of the GCs and galaxies in  
Table\,1, taken from NED, van den Bergh 2000, Harris 1996, McLaughlin \& van 
der Marel 2005, and other works for the UCDs as given in the text).
Clearly, more luminous galaxies possess more luminous GCs/UCDs. Is this
just a sampling effect reflecting the ever richer globular cluster systems?

\begin{figure}[t!]
\epsfysize=10.0cm
\epsfbox{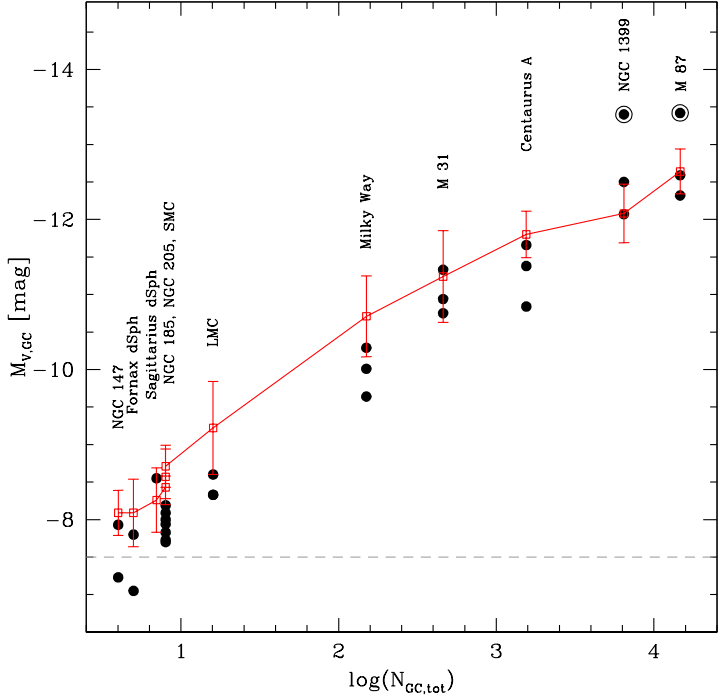}
\vspace*{-0.2cm}
\caption{The absolute magnitude of the brightest two or three GCs/UCDs of
a galaxy as a function of total number of GCs belonging to the host galaxy.
The open squares with errobars indicate the average luminosity of the
brightest GC from Monte Carlo simulations of 10.000 GCLFs of the
respective galaxies. The two brightest UCDs in Fornax and Virgo (encircled
dots) have extended low surface brightness envelopes. The dashed line marks
the universal luminosity of the GCLF turnover magnitude.
}
\end{figure}

Many studies of the globular cluster luminosity function (GCLF, number of GCs
vs. magnitude) have shown that the bright end shape can be well described by 
a Gaussian with a universal turnover magnitude at $M_V=-7.5$ mag (see Richtler
2003 and references therein). The dispersion of the GCLF, $\sigma_{\rm GCLF}$, 
ranges from 0.8 to 1.3 mag and increases with increasing host galaxy 
luminosity (Jord\'an et al. 2007).
To test the hypothesis that the brightest GCs are statistically compatible
with a Gaussian GCLF, we determined the average luminosity of the brightest
GC from Monte Carlo simulations of 10.000 GCLFs of our sample galaxies.
The GCLF function is defined by the total number of GCs, $N_{\rm GC,tot}$ and
its width $\sigma_{\rm GCLF}$ (see Table\,1). In Fig.\,3 the results of those 
simulations (open squares with errorbars) are shown together with the
brightest GCs. With the exception of the brightest UCD in the Fornax and Virgo
cluster (encircled dots), the brightest GCs/UCDs of all galaxies are 
compatible with being drawn from a Gaussian GCLF.
This is at odds with what one would expect if UCDs were a distinct kind of
objects (as discussed in the previous section). Also there is no hint for a 
maximum luminosity of a GC/UCD. The absolute magnitudes of the brightest GCs 
linearly increase with the logarithm of $N_{\rm GC,tot}$ (see also Billett et 
al. 2002, Weidner et al. 2004).
At first glance, these findings might pose a problem for the 
hierarchical assembly of the most massive galaxies. If a central cluster 
galaxy like NGC\,1399 is the result of a merger of several $L^\ast$ or Milky 
Way-type galaxies, one would expect the brightest GCs of the resulting merger 
to have a luminosity of about $\omega$\, Cen. On the other hand, just during
those mergers the most massive GCs/UCDs might have formed. I come back to this
point in the next section.

\begin{figure}[t!]
\epsfysize=10.0cm
\epsfbox{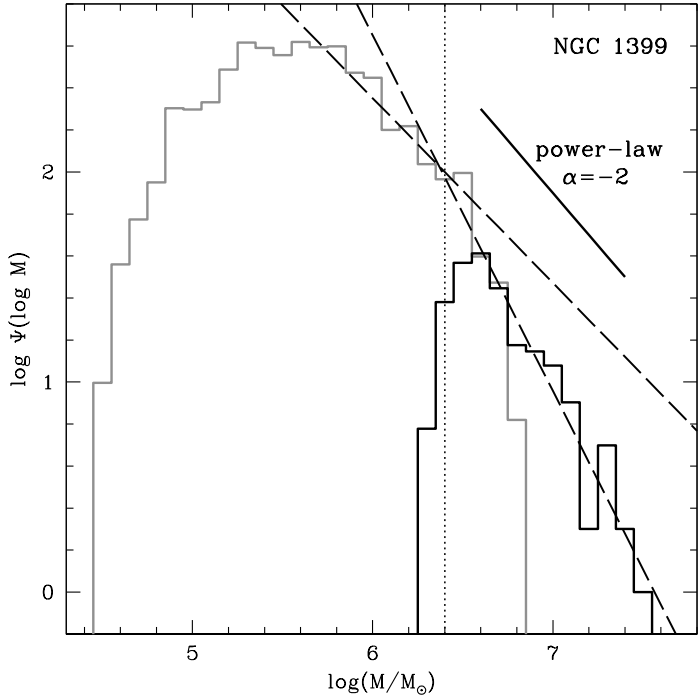}
\vspace*{-0.2cm}
\caption{Mass function of GCs and UCDs around NGC\,1399. The GCs (grey 
histogram) were taken from the Fornax ACS survey (Jord\'an et al. 2009). The 
black histogram is based on radial velocity members of the Fornax cluster (GCs
and UCDs, Hilker 2009, in prep.). The grey histogram was normalized to the
number counts of the black histogram at $\log M\simeq6.6 M_{\odot}$. The 
dashed lines are fits to the mass regimes $5.5<\log M<6.4$ and $6.6<\log 
M<7.5$ with power-law slopes $\alpha$ of $-1.9$ and $-2.7$, respectively. The 
dotted vertical line indicates the characteristic transition mass of 
$M_c=2.5\times10^6 M_{\odot}$ between GCs and UCDs.
}
\end{figure}

Before that, let us have a look at the mass function of GCs and UCDs in the
central Fornax cluster. The GCS of NGC\,1399 has the most complete coverage
of confirmed radial velocity members at the bright end of the GCLF, thanks to
massive spectroscopic surveys (Drinkwater et al. 2000, Richtler et al. 2004,
Mieske et al. 2004, Firth et al. 2007). More than 150 GCs/UCDs brighter than
$\omega$\,Cen are known (Hilker 2009, in prep.). The bulk of the lower mass 
GCs is well defined through the Fornax ACS survey (Jord\'an et al. 2009). 
Both datasets combined have been used to construct the mass function of GCs
and UCDs around NGC\,1399. First, the $gz$ photometry of the ACS data were
transformed into the Johnson $V$,$(V-I)$ system using the relation of Peng
et al. (2006, see also the CMD in Fig.\,1). Second, the mass-to-light ratio, 
$M/L_V$, of each GC/UCD was derived from its $(V-I)$ colour, using a fit to 
the $(V-I)$ and $M/L_V$ values of a 13-Gyr old single stellar population model
by Maraston (2005). A Kroupa IMF and a blue horizontal branch was assumed (see 
also Dabringhausen et al. 2008). $M/L_V$ and $M_V$, finally, were used to 
compute the masses of the GCs and UCDs.

\begin{figure}[t!]
\epsfysize=10.0cm
\epsfbox{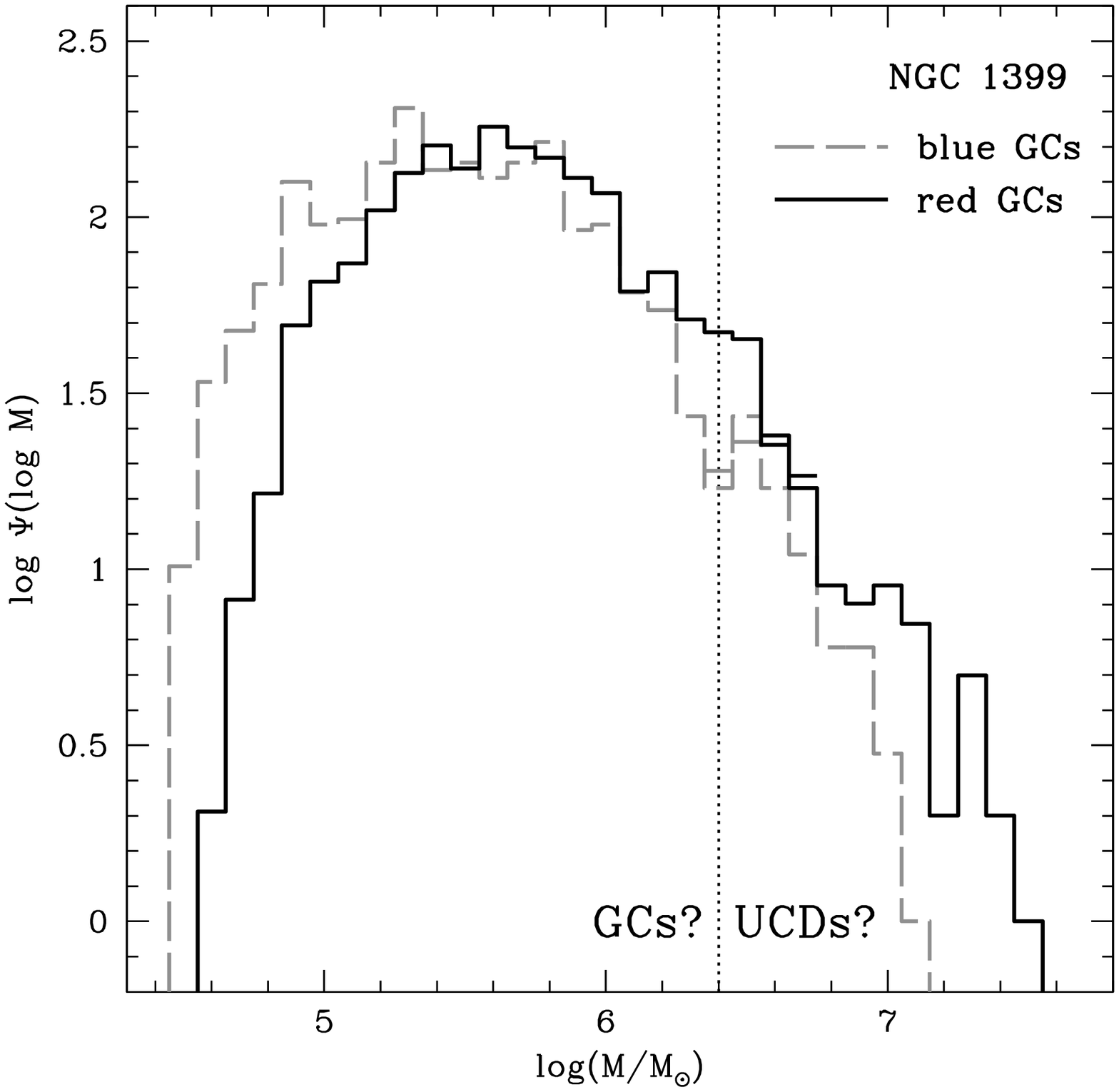}
\vspace*{-0.2cm}
\caption{Mass function of GCs and UCDs around NGC\,1399 (see Fig.\,4),
separated into blue (metal-poor) and red (metal-rich) GCs/UCDs as indicated.
The histograms of the GCs were normalized to the number counts of the
confirmed Fornax members at $\log M\simeq6.5 M_{\odot}$ and $\log M\simeq6.7 
M_{\odot}$ for the blue and red  GCs, respectively. The dotted vertical line
indicates the characteristic transition mass of $M=2.5\times10^6 M_{\odot}$
between GCs and UCDs.
}
\end{figure}

In Fig.\,4 the mass function of both samples is shown. The number counts of 
the ACS data were normalized to those of the spectroscopic sample in the
mass range $6.5<\log M<6.8 M_{\odot}$, a regime where both datasets are 
expected to be complete. The turnover magnitude $M_V=-7.5$ mag corresponds to
$\log M\simeq5.4$ which forms a plateau in the mass function. For masses
larger than $\log M>5.8$ the number counts are decreasing, but not with a 
uniform slope. In the mass range $5.5<\log M<6.4$ a fit to the data gives a 
power-law slope of $\alpha=-1.88$ (from $dN/dM\propto M^{-\alpha}$) which
also was found for other GCSs (e.g. Harris \& Pudritz 1994, Larsen et al. 
2001) and which is close to $\alpha=-2$, the typical slope for the mass 
functions of young cluster in merger galaxies (e.g. Zhang \& Fall 1999) 
and giant molecular clouds (e.g. Elmegreen 2002 and references therein). 
Beyond $\log M>6.5$ the mass function
falls off steeply. A fit to the data gives a slope of $\alpha=-2.70$.
Interestingly, both fits cross at $\log M\simeq6.4$, just the 
characteristic mass where the properties and scaling relations between GCs and
UCDs change ($M_c=2.5\times10^6 M_{\odot}$). Maybe there is some kind of
cut-off mass for `normal' GCs, and UCDs indeed follow a different formation 
mechanism?! Such a cut-off at the high mass end of the mass function was also
observed for young star clusters systems in spirals (e.g. Gieles et al. 2006),
although at an order of magnitude lower mass (Schechter 
function cut-off mass: $M_c=2.1\times10^5 M_{\odot}$, Larsen 2009).
For early-type galaxies in the Virgo cluster, Jord\'an et al. (2007) describe 
the GC mass function by an ``evolved Schechter function'' and show that
$M_c$ increases from $3\times10^5 M_{\odot}$ in bright dwarf
ellipticals ($M_V=-16$) to 2-3$\times10^5 M_{\odot}$ in giant 
ellipticals, consistent with what is presented here.

Fig.\,5 illustrates that the high mass end of GCs/UCDs is dominated by
metal-rich objects. As a division between blue (metal-poor) and red 
(metal-rich) GCs/UCDs a colour of $(V-I)=1.05$ mag ([Fe/H]$\simeq -0.8$ dex)
was chosen (see Fig.\,1). This colour corresponds to the well known dip in 
the bimodal colour distribution of GCs in elliptical galaxies (e.g. Gebhardt 
\& Kissler-Patig 1999).

\section{Formation scenarios for UCDs}

Various formation scenarios have been suggested to explain the origin of UCDs.
The three most promising and their implications concerning the presented 
properties of UCDs are:

\vskip1mm

\noindent
{\bf 1)} UCDs are the remnant nuclei of galaxies that have been significantly 
stripped in the cluster environment (e.g. Bassino et al. 1994, Bekki et al. 
2001). Numerical simulations have shown that nucleated dEs can be disrupted in
a galaxy cluster potential under specific conditions and that the remnant 
nuclei resemble UCDs in their structural parameters (Bekki et al. 2003) and 
mass-to-light ratio (Goerdt et al. 2008). 
In Fornax and Virgo, the small number of UCDs in both clusters points to a 
rather selective ``threshing'' process. The high metallicity of most Fornax 
UCDs seems to disfavour this scenario for their origin, whereas the brightest,
metal-poor GCs/UCDs indeed share most of the properties of present-day nuclei.
Note that that the threshing process also seems to work in our Galaxy. Good 
candidates for (former) nuclei are $\omega$ Cen (e.g. Hilker \& Richtler 2000)
and M54, the nuclear cluster of the Sagittarius dSph (e.g. Monaco et al. 
2005).

\vskip1mm

\noindent
{\bf 2)} UCDs have formed from the agglomeration of many young, massive star
clusters that were created during merger events (e.g. Kroupa 1998, Fellhauer
\& Kroupa 2002), like the Antennae galaxies where
many young super-star cluster complexes were found (e.g. Whitmore et al. 
1999). An evolved example of such a merged star cluster complex might be the 
300 Myr old, super-star cluster W3 in NGC 7252 (Maraston et al. 2004, 
Fellhauer \& Kroupa 2005). Indeed, a further passive evolution of W3 would
bring it into the regime of the most massive, metal-rich UCDs.
Moreover, the young massive star clusters in starburst/merger galaxies follow
a mass-size relation that is consistent with that of UCDs (Kissler-Patig et 
al. 2006).
If the old UCDs in Fornax and Virgo formed like this, the galaxy mergers
must have happened early in the galaxy cluster formation history when the 
merging galaxies were still gas-rich. However, these early mergers must 
have already possessed close to solar metallicity gas or they were 
self-enriched fast. Moreover, the stellar mass
function of the young star clusters must have been non-canonical to explain
the elevated $M/L$ values of UCDs. The small number of UCDs would imply that 
only the most massive star cluster complexes survived as bound systems (e.g.
Bastian et al. 2006).

\vskip1mm

\noindent
{\bf 3)} UCDs are the brightest globular clusters and were formed in the same
GC formation event as their less massive counterparts (e.g. Mieske et al. 
2004). The smooth shape of the bright end of the GC luminosity function (no 
excess objects!) might support this scenario.
The most massive GCs then supposedly formed from the most massive molecular
clouds (MCs) of their host galaxy, assuming that more massive galaxies (like 
M87) were able to form higher mass MCs than lower mass galaxies (like M31). 
The luminosity-size relation of the most massive clusters suggests that there
is a break of the formation/collapse physics at a critical MC mass. The high
$M/L$ values of the most massive GCs then would point either to a formation
of GCs in dark matter halos (e.g. Baumgardt \& Mieske 2008 and references 
therein)
or to a non-canonical (probably top-heavy) IMF that accompanies the formation
of the most massive GCs (e.g. Murray 2009, Dabringhausen et al. 2009).

\vskip1mm

\noindent
{\bf 4)} UCDs are genuine compact dwarf galaxies, maybe successors of ancient
blue compact dwarf galaxies, that formed from small-scale peaks in the 
primordial dark matter power spectrum (Drinkwater et al. 2004).
This scenario has the advantage that no external processes, like mergers or
tidal disruption, are needed. However, due to the small numbers of UCDs,
this formation channel then seems to be a rare event and one might ask why no
compact galaxies with a mass inbetween UCD3 (in Fornax) and M32 have been 
found.

\vskip1mm

Which of these scenarios tells us the truth? Why is there a characterstic mass
at which the scaling relations and the slope of the mass function changes?

It is widely accepted that globular clusters are formed inside the cores of 
supergiant
molecular clouds (e.g. McLaughlin \& Pudritz 1996). The balance between 
coagulation and disruption processes of these cores shapes the GC mass 
spectrum. Up to a final cluster mass of $\sim 10^6 M_{\odot}$ this seems to 
be a well regulated scale-free process. Does the break in the GC mass function 
correspond to a maximum `allowed' molecular cloud mass from which a GC can 
form? If so, all GCs/UCDs above the corresponding `maximum' GC mass must have 
formed from the
coalescence of lower mass GCs (or proto-GCs). This can have happend on 
a very short timescale during the GC formation process itself or on a longer 
timescale via the merging of individual GCs either in a compact star cluster 
complex (e.g. Fellhauer \& Kroupa 2002) or through tidal friction in the core 
of a dwarf galaxy (e.g. Oh \& Lin 2000). Also, a
nucelar star cluster can grow via episodic star formation triggered by 
infalling gas in the centre of a gas-rich galaxy (e.g. Walcher et al. 2006).
Alternatively, if there does not exist a maximum `allowed' molecular cloud 
mass, the physics of the massive cluster formation within the MCs must be 
different than for lower mass GCs (see Murray 2009 for a possible solution). 

It is not up to this contribution to discuss which scenario is the most 
plausible one. Since UCDs come with different flavours (metal-poor vs.
metal-rich; with and without low surface brightness envelope; etc.)
they probably comprise a `mixed bag of objects' from different formation
channels.
 
\section{Conclusions and Outlook}

The most massive globular cluster of a galaxy scales with the luminosity
of the host galaxy and the richness of the globular cluster system. 
When taking a Gaussian function as representation of the bright end of the 
globular cluster luminosity function, no excess objects are needed to explain 
the most luminous GCs in their respective environments. This includes the 
so-called ``ultra-compact dwarf galaxies'' (UCDs) which were identified as the 
brightest compact ($R_{\rm eff}<100$ pc) objects in nearby galaxy clusters,
but also around individual galaxies. 
Although there seems to exist a smooth luminosity function between GCs and 
UCDs, the mass function shows a break at a 
characteristic mass of $M_c\simeq 2.5\times 10^6 M_\odot$. Whereas GCs in the 
mass range $3.0\times10^5<M<2.5\times 10^6 M_\odot$ follow a power-law slope 
of $\alpha\simeq-1.9$ consistent with the measured power spectrum of molecular
clouds and young star clusters, compact objects (GCs/UCDS) above
$M_c$ are not as abundant as `normal' GCs. The slope falls off with an exponent 
$\alpha\simeq-2.7$. Strikingly, this characteristic mass also marks the
change of some key properties between GCs and UCDs. The most remarkable
properties of UCDs are that their size scales with their luminosity and that
their dynamical mass-to-light ratio is on average twice that of
GCs at a given metallicity. Moreover, the most massive UCDs seems to be 
exclusively metal-rich.
Although many of these characteristics are consistent with the known scaling
relations and properties of early-type galaxies, there exists a prominent
gap bewteen the most massive UCDs and the M32-type galaxies, the latter being
$\sim15$ times more massive than UCDs. This makes it unlikely that UCDs are
pure genuine compact galaxies related to small-scale dark matter clumps. 
Rather they are connected to gas-dynamical 
cluster formation processes, either as nuclear star cluster of nowadays 
dissolved galaxies or as merged super-star clusters which formed in violent 
starbursts such as seen in merging galaxies.
The latter scenario is supported by the existence of young massive star 
clusters with similar masses and scaling relations as those of UCDs.
The elevated $M/L$ values of UCDs, however, suggests that they were born
with a different (probably top-heavy) initital mass function than lower mass
GCs.

While we have some ideas on the possible origin of UCDs, there are many 
questions left to answer concerning their nature. Some 
important ones are: Do UCDs have multiple stellar populations? Can we find
young or intermediate age UCDs in the local universe? Do the large $M/L$
values really point to unusual initial mass functions? Or do they contain
dark matter? Is there tidal structure around UCDs? Do UCDs harbour black 
holes? 

Some of these questions will be answered in the next years with the help of
ongoing and future observing programmes. The results will bring more light
into the nature of these enigmatic objects.

\subsection*{References}

{\small
\bref
Barmby, P., Huchra, J.\,P., Brodie, J.\,P. 2001, AJ 121, 1482

\bref
Bassino, L.\,P., Muzzio, J.\,C., Rabolli, M. 1994, ApJ 431, 634

\bref
Bastian, N., Emsellem, E., Kissler-Patig, M., Maraston, C. 2006, A\&A 445, 471

\bref
Baumgardt, H., Mieske, S. 2008, MNRAS 391, 942

\bref
Bedin, L.\,R., Piotto, G., Anderson, J., et al. 2004, ApJL 605, L125

\bref
Bekki, K., Couch, W.\,J., Drinkwater, M.\,J. 2001, ApJ 552, L105

\bref
Bekki, K., Couch, W.\,J., Drinkwater, M.\,J., Shioya, Y. 2003, MNRAS, 344, 399

\bref
Billett, O.\,H., Hunter, D.\,A., Elmegreen, B.\,G. 2002, AJ 123, 1454

\bref
Buonanno, R., Corsi, C.\,E., Castellani, M., et al. 1999, AJ 118, 1671 

\bref
Carraro, G., Zinn, R., Moni Bidin, C. 2007, A\&A 466, 181

\bref
Carraro, G. 2009, AJ, in press (arXiv:0901.2673)

\bref
C\^ot\'e, P., Piatek, S., Ferrarese, L., et al. 2006, ApJS 165, 57

\bref
Crowl, H.\,H., Sarajedini, A., Piatti, A.\,E., et al. 2001, AJ 122, 220

\bref
Dabringhausen, J., Hilker, M., Kroupa, P. 2008, MNRAS 386, 864

\bref
Dabringhausen, J., Kroupa, P., Baumgardt, H. 2009, MNRAS, in press 
(arXiv:0901.0915)

\bref
Da Costa, G.\,S., Mould, J.\,R. 1988, ApJ 334, 159

\bref
Drinkwater, M.\,J., Gregg, M.\,D., Couch, W.\,J., et al. 2004, PASA 21, 375

\bref
Drinkwater, M.\,J., Gregg, M.\,D., Hilker, M., et al. 2003, Nature 423, 519

\bref
Drinkwater, M.\,J., Jones, J.\,B., Gregg, M.\,D., Phillipps S. 2000, PASA 17,
227

\bref
Elmegreen, B.\,G. 2002, ApJ 564, 773

\bref
Evstigneeva, E.\,A., Drinkwater, M.\,J., Peng, C.\,Y., et al. 2008, AJ 136, 
461

\bref
Evstigneeva, E.\,A., Gregg, M.\,D., Drinkwater, M.\,J., Hilker, M. 2007, AJ 
133, 1722

\bref
Fellhauer, M., Kroupa, P. 2002, MNRAS 330, 642

\bref
Fellhauer, M., Kroupa, P. 2005, MNRAS 359, 223

\bref
Firth, P., Drinkwater, M.\,J., Evstigneeva, E.\,A. 2007, MNRAS 382, 1342

\bref
Gebhardt, K., Kissler-Patig, M. 1999, AJ 118, 1526

\bref
Gieles, M., Larsen, S.\,S., Bastian, N., Stein, I.\,T. 2006, A\&A 450, 129

\bref
Glatt, K., Grebel, E.\,K., Sabbi, E., et al. 2008, AJ 136, 1703

\bref
Goerdt, T., Moore, B., Kazantzidis, S., et al. 2008, MNRAS 385, 2136

\bref
Harris, W.\,E. 1996, AJ 112, 1487

\bref
Harris, W.\,E., Pudritz, R. 1994, ApJ 492, 177

\bref
Ha\c{s}egan, M., Jord\'an, A., C\^ot\'e, P., et al. 2005, ApJ 627, 203

\bref
Hau, G.\,K.\,T., Spitler, L.\,R., Forbes, D.\,A. 2009, MNRAS, in press 
(arXiv:0901.1693)

\bref
Hilker, M. 1998, PhD thesis, Sternw.~Bonn, (1998)

\bref
Hilker, M., Baumgardt, H., Infante, L., et al. 2007, A\&A 463, 119

\bref
Hilker, M., Infante, L., Vieira, G., et al. 1999, A\&AS 134, 75

\bref
Hilker, M., Richtler, T. 2000, A\&A 362, 895

\bref
Hodge, P.\,W. 1973, ApJ 182, 671

\bref
Hodge, P.\,W. 1974, PASP 86, 289

\bref
Hodge, P.\,W. 1976, AJ 81, 25

\bref
Jones, J.\,B., Drinkwater, M.\,J., Jurek, R., et al. 2006, AJ 131, 312

\bref
Jord\'an, A., Peng, E.\,W., Blakeslee, J.\,P., et al. 2009, ApJS 180, 54

\bref
Jord\'an, A., McLaughlin, D.\,E., C\^ot\'e, P. 2007, ApJS 171, 101

\bref
Kissler-Patig, M., Jord\'an, A., Bastian, N. 2006, A\&A 448, 1031

\bref
Koch, A., Grebel, E.\,K., Wyse, R.\,F.\,G., et al. 2007, AJ 131, 895

\bref
Kroupa, P. 1998, MNRAS 300, 200

\bref
Kroupa, P. 2001, MNRAS 322, 231

\bref
Larsen, S.\,S. 2009, A\&A 494, 539

\bref
Larsen, S.\,S., Brodie, J.\,P., Huchra, J.\,P., et al. 2001, AJ 121, 2974

\bref
Mackey, A.\,D., Gilmore, G.\,F. 2003, MNRAS 345, 747

\bref
Mackey, A.\,D., Gilmore, G.\,F. 2004, MNRAS 352, 153

\bref
Maraston, C. 2005, MNRAS 362, 799

\bref
Maraston, C., Bastian, N., Saglia, R.\,P., et al. 2004, A\&A 416, 467

\bref
McLaughlin, D.\,E., Pudritz, R.\,E. 1996, ApJ 457, 578

\bref
McLaughlin, D.\,E., van der Marel, R.\,P. 2005, ApJS 161, 304

\bref
Meylan, G., Sarajedini, A., Jablonka, P., et al. 2001, AJ 122, 830

\bref
Mieske S., Kroupa, P. 2008, ApJ 677, 276

\bref
Mieske S., Hilker, M., Infante, L. 2004, A\&A 418, 445

\bref
Mieske, S., Hilker, M., Infante, L., Jord\'an, A. 2006, AJ 131, 2442

\bref
Mieske, S., Hilker, M., Infante, L., Mendes de Oliveira, C. 2007b, A\&A 463,
503

\bref
Mieske, S., Hilker, M., Jord\'an, A., et al. 2007a, A\&A 472, 111

\bref
Mieske, S., Hilker, M., Jord\'an, et al. 2008, A\&A 487, 921

\bref
Milone, A.\,P., Bedin, L.\,R., Piotto, G., et al. 2008, ApJ 673, 241

\bref
Minniti, D., Kissler-Patig, M., Goudfrooij, P., Meylan, G. 1998, AJ 115, 121

\bref
Misgeld, I., Mieske, S., Hilker, M. 2008, A\&A 486, 697

\bref
Monaco, L., Bellazzini, M., Ferraro, F.\,R., Pancino, E. 2005, MNRAS 356, 1396

\bref
Murray, N. 2009, ApJ 691, 946

\bref
Oh, K.\,.S., Lin, D.\,N.\,C. 2000, ApJ 543, 620

\bref
Peng, E.\,W., Ford, H.\,C., Freeman, K.\,C. 2004, ApJS 150, 367

\bref
Peng, E.\,W., Jord\'an, A., C\^ot\'e, P., et al. 2006, ApJ 639, 95

\bref
Phillipps, S., Drinkwater, M.\,J., Gregg, M.\,D., Jones, J.\,B. 2001, ApJ 560,
201

\bref
Piotto, G., Bedin, L.\,R., Anderson, J., et al. 2007, ApJL 661, L53

\bref
Rejkuba, M., Dubath, P., Minniti, D., Meylan, G. 2007, A\&A 469, 147

\bref
Richtler, T. 2003, `The Globular Cluster Luminosity Function: New Progress 
in Understanding an Old Distance Indicator', in: Stellar Candles for the 
Extragalactic Distance Scale, Lecture Notes in Physics, Berlin Springer 
Verlag, vol.\,635, p.281

\bref
Richtler, T., Dirsch, B., Gebhardt, K., et al. 2004, AJ, 127, 2094

\bref
Sollima, A., Pancino, E., Ferraro, F.\,R., et al. 2005, ApJ 634, 332

\bref
van den Bergh, S. 2000, The Galaxies of the Local Group, published by 
Cambridge, UK: Cambridge Univ. Press, Cambridge

\bref
van de Ven, G., van den Bosch, R.\,C.\,E., Verolme, E.\,K., de Zeeuw, P.\,T.
2006, A\&A 445, 513

\bref
Villanova, S., Piotto, G., King, I.\,R., et al. 2007, ApJ 663, 296

\bref
Walcher, C.\,J., van der Marel, R.\,P., McLaughlin, D., et al. 2006, ApJ 618,
237

\bref
Wehner, E., Harris, W.\,E. 2008, ApJL 668, 35

\bref
Weidner, C., Kroupa, P., Larsen, S.\,S. 2004, MNRAS 350, 1503

\bref
Whitmore, B.\,C., Zhang, Q., Leitherer, C., et al. 1999, AJ 118, 1551

\bref
Zhang, Q., Fall,S.\,M. 1999, ApJ 527, L81
}

\vfill

\end{document}